\documentclass[%
 aip,
 jmp,%
 amsmath,amssymb,
 reprint,%
 superscriptaddress
]{revtex4-1}

\usepackage{graphicx}
\usepackage{dcolumn}
\usepackage{bm}
\usepackage{multirow}

\usepackage{titlesec}
\titlespacing*{\subsection}
{0pt}{1.5ex}{0.5ex}
\titlespacing*{\section}
{0pt}{2.5ex}{1.0ex}

\begin{document}


\title{Automated Identification of Relevant Frontier Orbitals for Chemical Compounds and Processes}

\author{Christopher J. Stein}
\affiliation{ 
ETH Z\"urich, Laboratorium f\"ur Physikalische Chemie, Vladimir-Prelog-Weg 2, 8093 Z\"urich, Switzerland
}
\author{Markus Reiher}
\email[Corresponding author: ]{markus.reiher@phys.chem.ethz.ch}
\affiliation{ 
ETH Z\"urich, Laboratorium f\"ur Physikalische Chemie, Vladimir-Prelog-Weg 2, 8093 Z\"urich, Switzerland
}

\date{\today}

\begin{abstract}
Quantum-chemical multi-configurational methods are required for a proper description of static electron correlation, a phenomenon inherent to the electronic structure of molecules with multiple (near-)degenerate frontier orbitals.
Here, we review how a property of these frontier orbitals, namely the entanglement entropy is related to static electron correlation.
A subset of orbitals, the so-called \textit{active orbital space}, is an essential ingredient for all multi-configurational methods.
We proposed an automated selection of this active orbital space, that would otherwise be a tedious and error prone manual procedure, based on entanglement measures.
Here, we extend this scheme to demonstrate its capability for the selection of consistent active spaces for several excited states and along reaction coordinates.
\end{abstract}

\maketitle

\setlength{\parindent}{0cm}
\setlength{\parskip}{0.6em plus0.2em minus0.1em}

\section{Introduction}
Quantum chemistry is primarily concerned with the calculation of electronic energies of molecules and materials, including their properties which can be expressed as derivatives of the energy. 
Such an energy is calculated from the electronic wave function or the electron density.
A plethora of methods has been developed that allows chemists to determine the electronic wave function, the electron density, and the corresponding electronic energy. .
These methods typically scale unfavourably with system size, so that approaches which yield remarkably accurate results for small molecules\cite{kart06,klop10,fell12} are not feasible for large molecules.
It is this trade-off between accuracy and feasibility that makes the selection of an appropriate method a central task in every quantum chemical investigation and
stimulates the development of new methods.
In addition to the size of a molecule, other factors make an electronic structure calculation a potentially difficult task.

\begin{figure}
\includegraphics[width=0.5\textwidth]{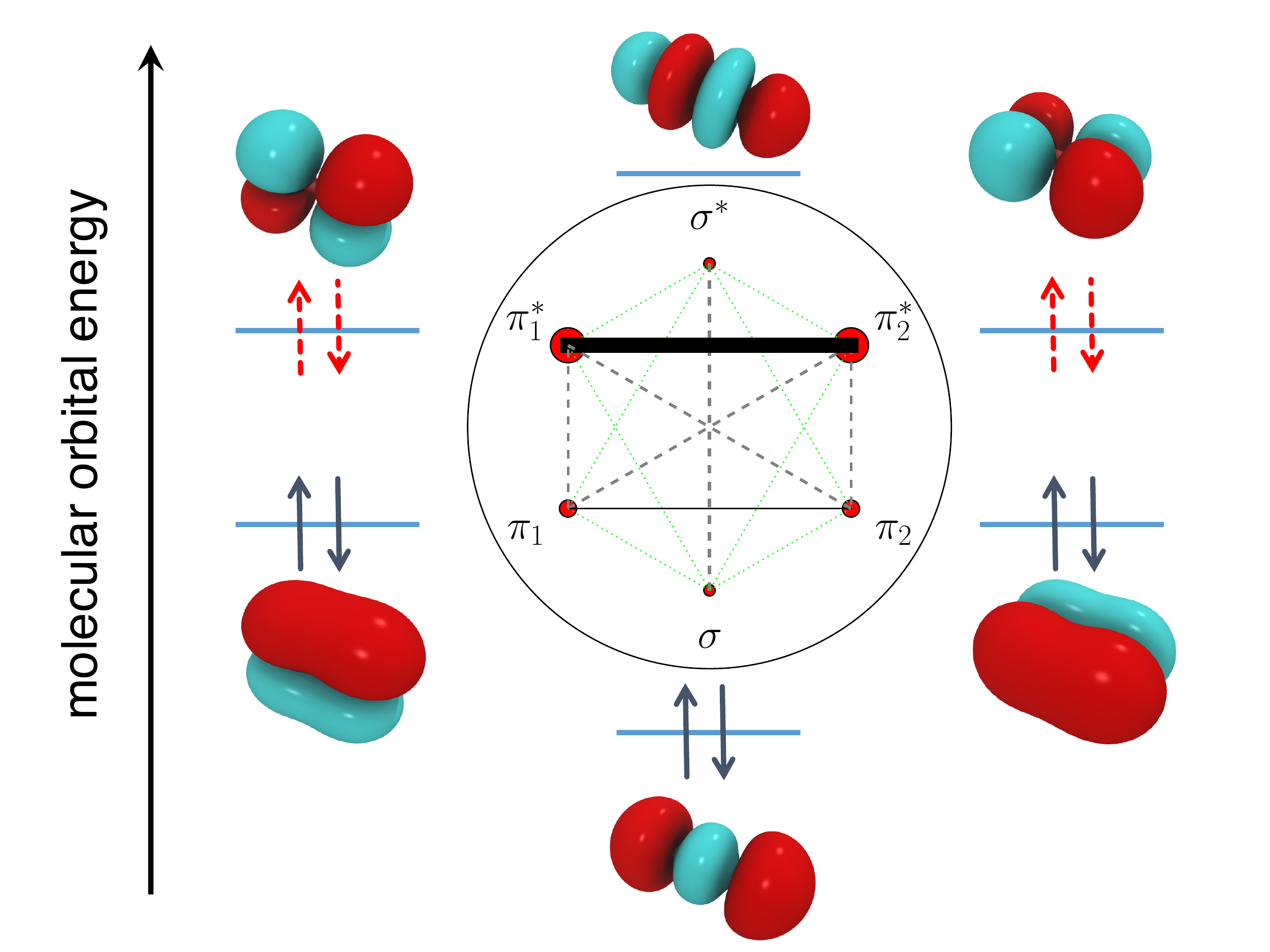}
\caption{Part of a molecular orbital diagram for the frontier orbitals of the oxygen molecule in its lowest singlet state ($^1\Delta_g$) consisting of eight electrons in six orbitals. The dark blue arrows denote electrons that are distributed over the orbitals according to the Aufbau principle. The fact that the electron pair is equally likely to be in either of the $\pi^*$-orbitals is indicated by the red dashed arrows. The corresponding entanglement diagram (see text) is included in the center. Here, every orbital is assigned a red circle whose area is proportional to its single-orbital entropy. The connecting lines mirror the value of the mutual information with the green dashed lines indicating a low, the gray dashed lines an intermediate, and the black solid lines a high value of the mutual information.}
\label{mo_diagram}
\end{figure}

One of these potential difficulties is known as \textit{static electron correlation}, emerging from dense-lying frontier orbitals. 
It can be illustrated at the simple example of the oxygen molecule in its lowest singlet spin state.
The electronic wave function is usually expanded in antisymmetrized products of molecular orbitals or symmetry-adapted linear combinations thereof.
In Figure~\ref{mo_diagram}, we show the frontier molecular orbitals of dioxygen.
Six of the eight electrons occupy the bonding $\sigma$- and $\pi$-orbitals according to the Aufbau principle.
The remaining electron pair must then occupy one of the degenerate antibonding $\pi^*$-orbitals, but none can be preferred over the other.
Any possible distribution of the electrons over the orbitals is referred to as an electronic \textit{configuration}. 
For singlet dioxygen in its $^1\Delta_g$ state there exist two electronic configurations of which one cannot be preferred over the other.
Note that the electronic configurations where two electrons of opposite spin singly occupy the degenerate $\pi^*$-orbitals give rise to a $^1\Sigma_g^+$ state of higher energy.

To describe the electronic structure of the  $^1\Delta_g$ state qualitatively correct, both configurations have to be included in a qualitatively correct approximation of the wave function.
The necessity to include more than one configuration represents the static-correlation problem and is encountered when molecular frontier orbitals are degenerate or near-degenerate.
It is therefore prototypical for the electronic structure of many transition-metal complexes and clusters, molecules with extended $\pi$-systems, and for chemical bonds that are broken or formed.

Many of the commonly applied quantum-chemical methods such as Kohn-Sham density functional theory (DFT)\cite{parr94} and coupled-cluster methods\cite{helg14} are based on reference wave functions comprising a single configuration only .
This deficiency of the reference wave function can often be corrected for multi-configurational systems.
Where appropriate, high-excitation coupled-cluster expansions or even broken-spin-symmetry approaches may work.
However, the only natural way to deal with the static correlation problem is to apply an intrinsically multi-configurational method.

The \textit{de facto} standard method for the calculation of strongly correlated molecules in quantum chemistry is the complete active space self-consistent field (CASSCF) method.\cite{roos80,wern85,roos87,shep87}
It implements the concept of a fully optimized reaction space\cite{rued76,rued82,rued82a,rued82b} and includes all configurations that can arise from a selected subset of so-called \textit{active} orbitals $L$ and electrons $N$, while the occupations of the low-energy \textit{inactive} orbitals and the virtual orbitals are fixed to either doubly occupied or unoccupied, respectively. 
The \textit{active space}, often denoted as CAS($N$,$L$), is usually manually selected from frontier orbitals that are likely to be highly statically correlated (in our example the $\pi^*$-orbitals of dioxygen in its $^1\Delta_g$ state).
This manual selection, however, is an error prone and tedious task,\cite{bach06,malm08,very11} especially for molecules with a large number of frontier orbitals such as polynuclear transition-metal clusters.
In addition, the scaling of these methods is exponential with the size of the active space such that active spaces with more than 18 electrons in 18 orbitals are not accessible to traditional CASSCF.\cite{aqui8}

New methods overcome this unfavourable scaling of the traditional approach.
The density matrix renormalization group (DMRG)\cite{whit92,whit93,lege08,chan08,chan09,mart10,mart11,chan11,scho11,kura14,wout14,yana15,szal15,knec16,chan16} and full configuration-interaction quantum Monte-Carlo\cite{boot09,boot10} are two examples.
Although these approaches allow for active spaces of 50-100 orbitals, the need for an active space selection remains important for two reasons:
\begin{enumerate}
\item The inclusion of only weakly statically correlated orbitals blows up the calculation and ultimately renders calculations on large molecules with many frontier orbitals impossible.
\item Methods that capture the missing part of electron correlation (so-called \textit{dynamical} correlation) may fail for unbalanced active spaces or whenever an undefined but considerably large fraction of this dynamical correlation is already represented in the active space.
\end{enumerate}

The selection of a meaningful active space requires experience with both multi-configurational methods and the system under investigation.
A scheme where the selection of the active orbitals is based on unbiased physical quantities is therefore highly desirable and ultimately enables the full automation of the active space selection.
The natural orbital occupation numbers (NOONs) of an unrestricted Hartree--Fock calculation\cite{pula88,bofi89,kell15a} or a M{\o}ller-Plesset perturbation theory calculation to second order (MP2)\cite{jens88} are suitable quantities for this purpose.
However, the small deviations from 0.00 or 2.00 that are typical for NOONs, especially for systems with a large number of close-lying frontier orbitals, make the definition of a suitable and general selection threshold a difficult task.
Therefore, we introduced an automated active orbital space selection protocol\cite{stei16,stei16a} based on orbital entanglement measures that are closely related to static correlation effects.\cite{bogu12}
Very recently, further work along these lines was presented.\cite{sayf17}

In this work, we briefly review the orbital entanglement quantities which are the basis of our protocol and discuss them at the dioxygen example in its $^1\Delta_g$ state.
For ethylene we then show how this approach is applied to describe several excited states of different spin symmetry.
Finally, we investigate the Diels-Alder reaction of ethylene with 1,3-butadiene to demonstrate how our protocol selects a consistent active space along a reaction coordinate.


\section{Orbital Entanglement and Review of Previous Work}

For a spatial orbital four \textit{pure states} can be defined: doubly occupied, spin up, spin down, and unoccupied.
A single electron configuration assigns each orbital one of these pure states unambiguously so that, in the light of the definition of static correlation, each orbital
contributes an (almost) pure sub-state in the absence of static correlation (i.e. one configuration is sufficient to approximate the wave function).

By contrast, when static correlation is pronounced as in the example of the $^1\Delta_g$ state of dioxygen (see Figure~\ref{mo_diagram}), the occupation of some of the orbitals will deviate substantially from a pure state.
In this example, the occupation of each of the $\pi^*$-orbitals is mainly a superposition of a doubly occupied and an empty single-orbital sub-state.
The deviation from such a pure state is precisely what necessitates the inclusion of several configurations in the reference wave function (equivalently,
it is the presence of several configurations with large weights in the qualitatively correct wave function that produces the deviation from a pure state).
A quantity that measures this deviation from a pure state and that can be obtained at low cost may then guide the selection of active orbitals.

The information entropy of a bipartite system is such a measure.\cite{lege03,riss06}
Defined for a single orbital $i$ it is referred to as the single-orbital entropy $s_i(1)$ and can be written as\cite{lege03,lege06}
\begin{align}
s_i(1) =- \sum_{\alpha=1}^4 w_{\alpha,i}\ln w_{\alpha,i},
\end{align}
where $\alpha$ runs over the four possible orbital occupations and $w_{\alpha,i}$ are the eigenvalues of the one-orbital reduced density matrix (1o-RDM) of orbital $i$, in which all environment sub-states defined on all other orbitals but $i$ were traced out.
It can be derived from the one- and two-particle reduced density matrices (1e- or 2e-RDM)\cite{riss06,bogu15} or as expectation values over strings of excitation and annihilation operators.\cite{bogu13}
Analogously, for a pair of orbitals the two-orbital entropy $s_{ij}(2)$ reads
\begin{align}
s_{ij}(2) =- \sum_{\alpha=1}^{16} w_{\alpha,ij}\ln w_{\alpha,ij},
\end{align}
where $\alpha$ includes now all combinations of the four one-orbital sub-states for each orbital and $w_{\alpha,ij}$ are the eigenvalues of the two-orbital reduced density matrix (2o-RDM) which now can be derived from RDM elements up to fourth order (4e-RDM).
Subtracting the individual single-orbital entropies from $s_{ij}(2)$ yields the \textit{mutual information} $I_{ij}$ and is a measure of the entanglement between sub-states defined on these two orbitals
\begin{align}
I_{ij} = \frac{1}{2}[s_i(1) + s_j(1) - s_{ij}(2)](1-\delta_{ij}),
\end{align}
where prefactor and signs correspond to those implemented in our DMRG program \textsc{QCMaquis}\cite{dolf14,kell15,kell16} and we note that other definitions exist.\cite{bogu12,lege06}

Entanglement diagrams that collect these measures in one figure for a given calculation provide intuitive insight into static correlation.
One example is shown in the center of Figure~\ref{mo_diagram}.
Each orbital is represented by a red circle whose area is proportional to its single-orbital entropy.
The thickness of the connecting lines is proportional to the mutual information defined for a pair of orbitals.
In the singlet dioxygen example, the two $\pi^*$-orbitals have the highest single-orbital entropy and a strong mutual information as we would expect from our difficulties in applying the Aufbau principle.
The mutual information further reveals a significant entanglement of the two $\pi$-orbitals as well as $\pi$-$\pi^*$- and $\sigma$-$\sigma^*$-entanglement.

These entanglement measures can be evaluated from partially converged, and therefore comparatively cheap, DMRG calculations that include a large number of possibly statically correlated frontier orbitals (ideally the whole valence orbital space for a reliable selection).\cite{stei16}
Based on such calculations, groups of orbitals with a high single-orbital entropy are automatically identified in our active orbital selection protocol after scaling to the highest $s_i(1)$ value found in the calculation under consideration.
The final and fully converged DMRG and CASSCF calculation is then carried out with these orbitals only.
For details concerning the selection protocol, we refer to the original publication in Ref.~\citenum{stei16}.
We showed that suitable active orbital spaces were selected for many molecules that are known to be challenging and for which reliable active spaces were established in the literature.

At the example of homolytic dissociation energies of $3d$-metallocenes, we further demonstrated\cite{stei16a} that active spaces selected by our protocol are well-suited for the subsequent inclusion of dynamical correlation, which was in this case included by means of CASSCF perturbation theory to second order.\cite{ande92}
In addition, we defined a diagnostic for the multi-configurational character of a wave function $Z_{s(1)}$.\cite{stei16b}

\section{Methodology}
\begin{figure}
\includegraphics[width=0.5\textwidth]{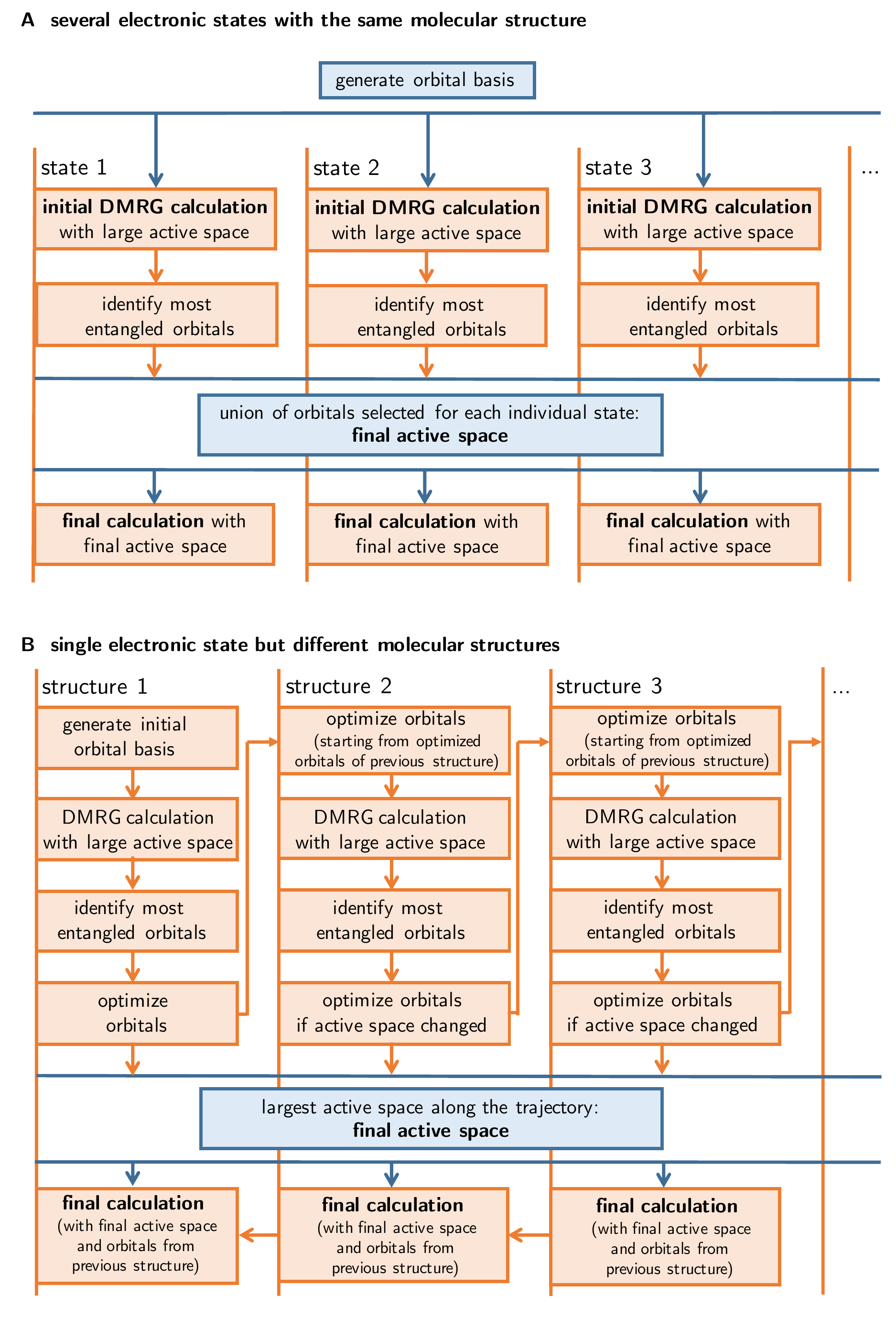}
\caption{Flowchart of the automated active space selection for: \textbf{A} several electronic states for a given molecular structure and \textbf{B} one electronic state but several molecular structures (e.g., along a dynamics trajectory or reaction coordinate). Steps in orange are carried out for an individual state/structure, while steps in blue combine information obtained from all states/structures or distribute information to all states/structures. Note that the direction along a reaction coordinate is in general not unique (it is possible to choose a coordinate from educts to products or vice versa) and the automated active space selection might suggest different final active spaces depending on the direction chosen. 
These differences, however, concern only orbitals with little entanglement entropy and their effect is levelled out by subsequent inclusion of dynamical correlation (see text).
This flowchart is an extension of the automated active space selection protocol proposed in Refs.~\citenum{stei16,stei16b}.}
\label{flowchart}
\end{figure}

In this work, we extend the automated active space selection to the calculation of transition energies for several excited singlet and triplet states of ethylene and to the Diels--Alder reaction of ethylene and \textit{cis}-butadiene.
The workflow is schematically shown in Figure~\ref{flowchart}.
The calculation of transition energies for several electronic states with an active space method requires that each state is calculated with the same orbital basis.
A state-specific orbital optimization in the final step might lead to slightly different shapes of the orbitals but the orbital character must remain the same for each orbital among all states to guarantee size-consistency of all calculations.
Therefore we first generate an orbital basis that is then applied in initial DMRG calculations with large active spaces for all states of interest (panel \textbf{A} of Figure~\ref{flowchart}).
This orbital basis can be Hartree--Fock or CASSCF orbitals of the ground state but other choices are possible, too.
With the automated active space selection protocol we select the most entangled orbitals in each state and construct the final active space from the union of these orbitals.
In a final step, the energies of all states are calculated from this active space and dynamical correlation is included.

When we aim to select a consistent active orbital space for several structures along a trajectory or reaction coordinate, we apply the workflow in panel \textbf{B} of Figure~\ref{flowchart}.
For a given initial structure we generate an initial orbital basis from which we extract the most entangled orbitals with our automated active space selection protocol and optimize these orbitals with state-averaged or state-specific CASSCF (or DMRG-SCF).
These orbitals are the input orbitals for the next structure, for which we optimize the orbitals again, select the highly entangled orbitals according to our protocol, and reoptimize the orbitals if the active space changed.
In this way the size of the active space can only \textit{increase} along the reaction coordinate because active orbitals with little entropy for some structures cannot be excluded from the calculation due to their importance for other structures.
After the procedure is applied for all structures and a final active space is identified, we go back to the structure, where the final active space was identified for the first time.
In order to obtain size-consistent energies and wave functions we now revert the direction and perform the final calculations with the final active space for each structure until we reach the first structure.
As the orbital optimization produces new orbitals for every structure, it is possible that the direction in which we traverse the reaction coordinate (i.e. educt to product or vice versa) affects the final active space.
We show, however, in the following that a) the neglect of orbitals with little entanglement may be leveled out by the inclusion of dynamical correlation and b) that a slight bias with respect to the entropies introduced by the orbital optimization of already selected active orbitals does not prevent the protocol from picking up highly entangled orbitals that are important to describe the reaction. 

All Hartree--Fock, CASSCF\cite{roos80,wern85} and CASPT2\cite{ande92} calculations were performed with \textsc{Molcas} 8\cite{aqui8}, while all DMRG calculations were carried out with our DMRG program \textsc{QCMaquis}.\cite{kell15,kell16,knec16}
We chose the ANO-RCC basis set\cite{pier95,roos04} in its double-zeta contraction for the study of the Diels--Alder reaction, whereas the standard contraction was applied for the calculation of the excited states of ethylene.
The active space selection was carried out with a newly developed graphical user interface that combines the automated active space selection protocol\cite{stei16}, with the protocols of Figure~\ref{flowchart} and additional analysis functionality such as the calculation of the multi-reference diagnostic $Z_{s(1)}$\cite{stei16b}.

\section{Singlet and Triplet Excited States of Ethylene}
The electronically excited states of ethylene were the topic of more than 60 theoretical articles and we refer to the reviews in Refs.~\citenum{peye75,bord96,cave97,buen99} and a recent article\cite{fell14} for a comprehensive overview.
Because of the size and symmetry of the molecule highly accurate approaches can be applied.
The surprisingly diffuse character of the $^1$B$_{1u}$ state is only captured by the most sophisticated methods.\cite{fell14}
Here, we illustrate benefits and possible drawbacks of our automated orbital selection approach.
We followed the protocol in panel \textbf{A} of Figure~\ref{flowchart} to calculate vertical transition energies of four excited singlet and triplet states of ethylene.
The initial set of orbitals for the DMRG calculations for all states included 24 (12 valence orbitals and 12 additional Rydberg orbitals) of the $^1$A$_g$ ground state Hartree--Fock orbitals.

\begin{table}
\caption{Vertical CASPT2 transition energies  $T_v$ (in eV) and $\langle x^2\rangle$ values (in $a_0^2$) for active spaces of three different sizes. Values in parentheses correspond to differences with respect to calculation with the largest active space.}
\hskip-0.4cm\begin{tabular}{l|rrrrrr|rr}
\hline \hline
 & \multicolumn{6}{c|}{} & ic- & \\
   & \multicolumn{6}{c|}{CASPT2 with orbitals from} & CAS-& \\
 &\multicolumn{2}{c}{auto-CAS(10,11)} & \multicolumn{2}{c}{CAS(12,16)} & \multicolumn{2}{c|}{CAS(12,19)}&CI$^a$&exp.$^b$\\
state & $T_v$ &$\langle x^2\rangle$ & $T_v$ & $\langle x^2\rangle$ & $T_v$ &$\langle x^2\rangle$ &$T_v$ & $T_v$\\
\hline
$^1$A$_g$      & 0.00           & 11.49 & 0.00 & 11.56 & 0.00 & 11.56 & 0.00 & 0.00 \\
$^1$B$_{3u}$ & 7.52 (0.06) & 20.98 & 7.46 & 21.05 & 7.46 & 20.96 & 7.39 & 7.11  \\
$^1$B$_{1u}$ & 8.31 (0.30) & 23.65 & 8.11 & 20.76 & 8.01 & 18.89 & 8.05 & 7.68  \\
$^1$B$_{1g}$ & 8.18 (0.04) & 18.53 & 8.15 & 18.83 & 8.14 & 18.75 & 7.98 & 7.8    \\
$^1$B$_{2g}$ & 8.20 (0.03) & 19.74 & 8.27 & 19.60 & 8.17 & 19.51 & 8.05 & 7.9    \\
$^3$B$_{3u}$ & 7.42 (0.07) & 20.94 & 7.35 & 20.98 & 7.35 & 20.95 & 7.24 & 6.98  \\
$^3$B$_{1u}$ & 4.53 (-0.01)& 11.45 & 4.55 & 11.64 & 4.54 & 11.64 & 4.54 & 4.36  \\
$^3$B$_{1g}$ & 8.14 (0.05) & 18.21 & 8.11 & 18.53 & 8.10 & 18.44 & 7.94 & 7.79  \\
$^3$B$_{2g}$ & 8.14 (0.04) & 19.32 & 8.20 & 19.32 & 8.10 & 19.00 & 7.97 & -       \\
\hline \hline
\multicolumn{9}{l}{$^a$Best estimate from Ref.~\citenum{fell14} internally contracted state-}\\
\multicolumn{9}{l}{average CAS configuration interaction calculations with core-}\\
\multicolumn{9}{l}{valence correction and extrapolation to the basis set limit of}\\
\multicolumn{9}{l}{hand-tailored atomic orbital bases with diffuse functions.}\\
\multicolumn{9}{l}{$^b$Values taken from Ref.~\citenum{fell14} with data from Ref.~\citenum{robi12}. }
\label{energies}
\end{tabular}
\end{table}

Table~\ref{energies} lists vertical transition energies $T_v$ and the expectation value $\langle x^2\rangle$ for the ground-state and four excited singlet and triplet states of ethylene, where $\langle x^2\rangle$ is a measure of the diffuseness of a state along the out-of-plane axis (the $x$ axis).
All entries of Table~\ref{energies} were calculated with CASPT2 with state-specific CASSCF orbitals for three different active spaces.
The first, and smallest, CAS(10,11) is the one selected by our automated procedure, whereas the two larger ones are taken from Ref.~\citenum{fell14}, the latter being the largest and recommended active space in that work.
A $\langle x^2\rangle$ value of around 11\,a$_0^2$ is typical for a pure valence state, one around $60$\,a$_0^2$ or higher is characteristic of a Rydberg state, and values in between correspond to mixed valence/Rydberg states.\cite{fell14}
A comparison with the best values from Ref.~\citenum{fell14} is not meaningful because the authors constructed a hand-tailored atomic orbital basis set with additional diffuse functions.
Moreover, their internally contracted state-average CAS configuration interaction method is more advanced than the CASPT2 approach we apply here for demonstration purposes.
We therefore compare our CASPT2 results with those obtained from a CASPT2 calculation with the CAS(12,19) that was recommended by the authors of Ref.~\citenum{fell14}.
It is noteworthy that the number of configurations is around 8\,000 in our case and around 18\,000\,000 for the large active space recommended in Ref.~\citenum{fell14}.
Nevertheless, our results are in very close agreement to those obtained with the large active space for both $T_v$ and $\langle x^2\rangle$ for all states but the problematic $^1$B$_{1u}$ state.
Hence, our method is capable of selecting the most entangled orbitals and therefore accurately accounts for the static correlation present in several electronic states.

An obvious explanation for the less accurate results that we obtained for the $^1$B$_{1u}$ state might be that the selection is based on orbitals that are optimized for the ground state and therefore differ substantially from the optimized orbitals for that particular state.
However, our automated protocol gives identical results when we apply the $^1$B$_{1u}$ CAS(12,19)SCF orbitals as the orbital basis for which we calculate the entropy measures.
This is not unexpected because this state is particularly sensitive to the method with which the dynamical correlation is calculated.\cite{fell14}
Even the highly-accurate internally contracted state-averaged CAS configuration interaction composite approach of Ref.~\citenum{fell14} had to be adapted for this specific state to give satisfactory results.
The results we obtained for this state are therefore not to be mistaken for a failure of our automated active orbital space selection because the static part of the electron correlation is adequately described.

\section{Concerted Diels--Alder Reaction of Ethylene with \textit{cis}-Butadiene}

The Diels--Alder reaction of ethylene and \textit{cis}-butadiene is a prototype for a [4+2] cycloaddition and was extensively studied with multi-configurational methods.\cite{bern85,li93,saka00,lisc04,chen14}
For this example we show that a consistent selection of active orbitals (and therefore a smooth potential energy surface) is possible with our automated approach (see panel \textbf{B} of Figure~\ref{flowchart}).
We chose a reaction coordinate consisting of 20 standard non-self-consistent density-functional tight-binding (DFTB)\cite{pore95,seif96} structures taken from Ref.~\citenum{haag14}.
These structures result from a constrained optimization along a $C_s$-symmetric reaction coordinate.
The large initial active orbital space for which we calculate the entanglement entropy contains all 34 valence orbitals (2$s$- and 2$p$-orbitals on carbon and 1$s$-orbitals on hydrogen) from a restricted Hartree--Fock calculation in case of the first structure.
The automatically selected active orbitals are then optimized with CASSCF and for all subsequent structures along the reaction coordinate, the optimized orbitals of the previous structure serve as starting orbitals for the CASSCF calculation.

\begin{figure}
\includegraphics[width=0.5\textwidth]{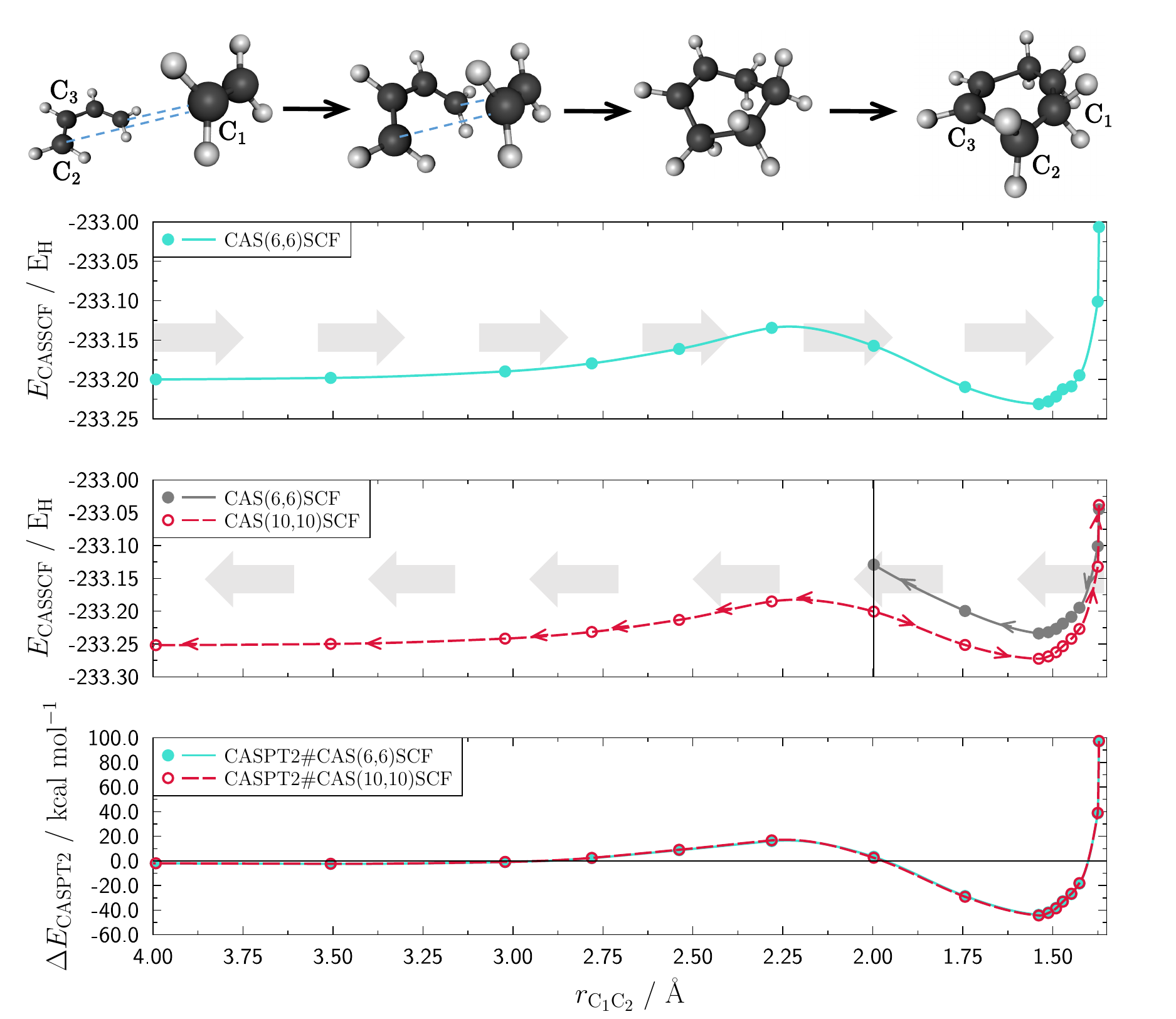}
\caption{Electronic energy profiles of the concerted Diels-Alder reaction of ethylene and \textit{cis}-butadiene. The reaction path is shown on the top.
The distance between the carbon atoms C$_1$ and C$_2$ is chosen as the reaction coordinate in a $C_s$-symmetric constrained optimization.
The upper energy diagram shows the CASSCF energy of a CAS(6,6) that is consistently selected by the automated active space selection for all structures considered when starting at the educts.
Defining the reaction coordinate in the opposite direction (middle panel) results in a CAS(6,6) (gray) until a C$_1$--C$_2$ distance of about 2.0~\AA is reached, where four additional orbitals are selected to yield a CAS(10,10) (red).
CASPT2 energies based on both CAS(6,6)SCF (turquoise) and CAS(10,10)SCF (red) calculations are shown in the lower panel and are virtually indistinguishable.}
\label{energy_scan}
\end{figure}

\begin{figure}
\includegraphics[width=0.5\textwidth]{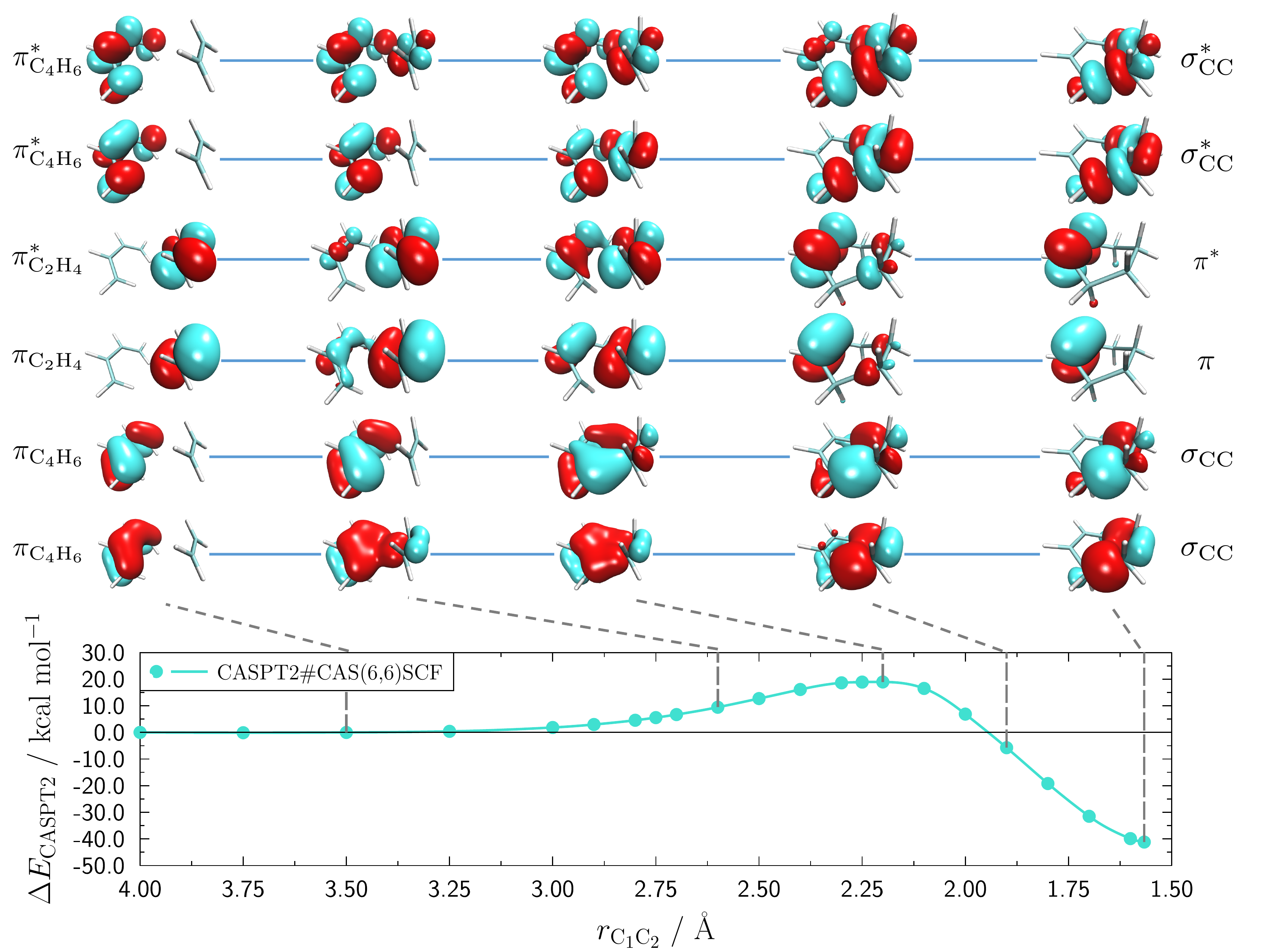}
\caption{Contour plots of the six active orbitals for five structures from a constrained CAS(6,6)SCF structure optimization along the $r_{\mathrm{C}_1\mathrm{C_2}}$ reaction coordinate of the concerted Diels--Alder reaction between ethylene and \textit{cis}-butadiene.
The lower part shows CASPT2 energies for the CAS(6,6)SCF structures and reference wave functions. 
The energies are taken relative to that of both molecules at infinite distance.}
\label{orbitals_scan}
\end{figure}

Results of these calculations are shown in Figure~\ref{energy_scan}.
When we define the reaction coordinate to start from separated molecules, the automatically selected active space is a CAS(6,6) for all structures along this coordinate (upper diagram of Figure~\ref{energy_scan}).
These active orbitals change their shape significantly during the reaction and resemble those displayed in the top part of Figure~\ref{orbitals_scan}.
While they describe the $\pi$-system of the separated molecules at early stages of the reaction, they transform to the two $\sigma_\mathrm{CC}$-bonds and the remaining $\pi/\pi^*$-orbitals of the product.
A transition state can be observed close to a C$_1$-C$_2$ distance of 2.25~\AA \,(see the upper part of Figure~\ref{energy_scan} for a definition of the reaction coordinate).
The most stable structure is found close to a C$_1$-C$_2$ distance of 1.55~\AA.
For even shorter C$_1$-C$_2$ distances, the electronic energy rises again and the C$_2$-C$_3$ distance increases significantly.

The energy diagram in the middle same Figure~\ref{energy_scan} shows the results of CASSCF calculations for automatically selected active spaces for the inverse definition of the reaction coordinate from right to left in Figure~\ref{energy_scan}.
Interestingly, the long C$_2$--C$_3$ distance of the initial structure is already indicating bond breaking as monitored by high orbital entropies for the $\sigma$- and $\sigma^*$-orbitals of these bonds in an Hartree--Fock basis.
Hence, the initially selected CAS(6,6) (gray circles in the central diagram of Figure~\ref{energy_scan}) contains those four orbitals as well as the $\pi/\pi^*$-orbitals of the cycloadduct.
When the C$_1$-C$_2$ distance reaches about 2~\AA, four additional orbitals are selected by the automated procedure (red circles in the central diagram of Figure~\ref{energy_scan}).
These orbitals are $\sigma$- and $\sigma^*$-orbitals of the bonds that are about to break.
This CAS(10,10) is also the largest active space selected along the reaction coordinate and is therefore our final active space.
Following the protocol of panel \textbf{B} of Figure~\ref{flowchart}, we apply this final active space now also for the remaining structures of the reaction coordinate (for which we had applied a smaller active space before) to obtain a smooth potential energy curve.

The discrepancy in the selected active spaces depending on the definition of the reaction coordinate may seem odd at first glance but has virtually no effect in actual calculations.
The first reason is that the thresholds in our automated procedure are designed to rather include orbitals with little entanglement than omitting important orbitals such that the resulting active spaces are guaranteed to include all static correlation effects.
This difficulty follows from the smooth transition between static and dynamic correlation.
The second reason is that in chemistry we are rarely interested in total energies and often aim to calculate relative energies.

In the lower diagram of Figure~\ref{energy_scan}, we present CASPT2 energies relative to the energy of the separated molecules for both active spaces.
Obviously, the energies are indistinguishable for both active spaces such that the missing correlation in the CAS(6,6)SCF calculation is fully recovered by the CASPT2 calculation.
This is in line with our previous work on the dissociation energy of metallocenes,\cite{stei16b} where missing size-consistency stemming from different sizes of the combined active spaces of the dissociated fragments and the complex was counterbalanced by the inclusion of dynamical correlation as calculated with CASPT2.
Therefore, our automated procedure will give consistent results as long as it guarantees that all static correlation effects are included in the reference wave function and hence that all \textit{strongly} entangled orbitals are included.
This is regulated by a single threshold in the automated protocol that is tuned accordingly.

For the final calculation of a reaction coordinate we focused on the region between the minimum in each of the diagrams of Figure~\ref{energy_scan} and the dissociated molecule.
In this region, only the six orbitals leading to the CAS(6,6) showed strong entanglement and were selected for CAS(6,6)SCF structure optimizations.
The minimum CAS(6,6)SCF structure was optimized without any constraint starting from the DFTB structure with $r_{\mathrm{C}_1\mathrm{C}_2} = 1.539$~\AA \, with a resulting  C$_1$-C$_2$ distance of 1.56643~\AA.
All other structures were calculated by constraining the C$_1$-C$_2$ distance (and C$_1$'-C$_2$' distance because of the $C_s$ symmetry) to 22 values between 1.6 and 4.0~\AA.
Corresponding CASPT2 energies relative to the energy of the dissociated molecule are displayed in Figure~\ref{orbitals_scan} along with contour plots of the six active orbitals for five structures along the reaction coordinate.
The potential energy curve is almost indistinguishable from the one obtained with the DFTB structures.
We note here that the change of the orbital character along this reaction coordinate might cause problems in the active space selection procedure if the changes between two subsequent structures are too pronounced so that large orbital rotations from the inactive or virtual space to the active orbital space occur during orbital optimization.
In that case, kinks in the CASSCF potential energy curve might occur and hence are a way to identify this possible obstacle.

\section{Conclusions and Outlook}
The selection of active orbital spaces is a tedious task that is essential to most multi-configurational calculations.
We reviewed our automated protocol that rates orbitals according to entanglement entropy measures.
This orbital entanglement is closely connected to the static correlation effects that are to be described by these multi-configurational calculations.\cite{bogu12}
While our previous work focused on static and ground-state properties, the emphasis here was on the consistent description of several excited states for a given molecular structure and the selection of an active space that is uniform along a reaction coordinate.
In both cases, the automated procedure identifies the union of orbitals that are selected for the individual states or structures as the final active space although slightly different protocols are applied.
We emphasize that our automated active space selection will not be possible whenever the large preliminary DMRG calculation is either unfeasible or not converged with respect to the CAS size.
The calculation of the orbital entanglement entropy facilitates the full automation of the automated active orbital selection procedure and ultimately enables non-experts to easily carry out multi-configurational calculations.

\section*{Acknowledgments}
This work was supported by the Schweizerischer Nationalfonds (No.~20020\_169120). 
C.J.S. gratefully acknowledges a K\'ekule fellowship from the Fonds der Chemischen Industrie.

\titlespacing*{\section}
{0pt}{1.5ex}{-2.0ex}
\section*{References}

%

\end{document}